\def\oP{\omega_{p_{0}}}
\def\bJ{{\beta_{_{J}}}}
\def\TsCO{T^{\star}_{CO}}

\def\Tp{{\textup{T}_p}}
\def\Pe{{P\left ( t^{esc} \right )}}

\def\wRA{{\omega _{_{dRA}}}}
\def\wRAa{{\omega _{_{dRA}}^{0.0}}}
\def\wRAb{{\omega _{_{dRA}}^{0.1}}}
\def\wRAc{{\omega _{_{dRA}}^{0.5}}}
\def\wRAd{{\omega _{_{dRA}}^{0.9}}}

\documentclass[aps,prb,twocolumn,groupedaddress,showkeys,superscriptaddress,floatfix]{revtex4-1}
\usepackage{epsfig}
\usepackage{multirow}
\usepackage{amsmath, amssymb}
\usepackage{color}
\usepackage{graphicx}
\usepackage{dcolumn}
\usepackage{bm}
\usepackage{subfigure}

\begin{document}

\bibliographystyle{apsrev4-1}

\title{\bf Phase dynamics in graphene-based Josephson junctions in the presence of thermal and correlated fluctuations}

\author{Claudio Guarcello\thanks{e-mail: claudio.guarcello@unipa.it}}
\affiliation{Group of Interdisciplinary Theoretical Physics Universit\`a di Palermo and CNISM, Unit\`a di Palermo Viale delle Scienze, Edificio 18, 90128 Palermo, Italy}
\affiliation{Radiophysics Department, Lobachevsky State University, 23 Gagarin Avenue, 603950 Nizhniy Novgorod, Russia}
\author{Davide Valenti\thanks{e-mail: davide.valenti@unipa.it}}
\affiliation{Group of Interdisciplinary Theoretical Physics Universit\`a di Palermo and CNISM, Unit\`a di Palermo Viale delle Scienze, Edificio 18, 90128 Palermo, Italy}
\author{Bernardo Spagnolo\thanks{e-mail: bernardo.spagnolo@unipa.it}}
\affiliation{Group of Interdisciplinary Theoretical Physics Universit\`a di Palermo and CNISM, Unit\`a di Palermo Viale delle Scienze, Edificio 18, 90128 Palermo, Italy}
\affiliation{Radiophysics Department, Lobachevsky State University, 23 Gagarin Avenue, 603950 Nizhniy Novgorod, Russia}
\affiliation{Istituto Nazionale di Fisica Nucleare, Sezione di Catania,  Via S. Sofia 64, I-90123 Catania, Italy}




\date{\today}

\begin{abstract}
In this work we study by numerical methods the
phase dynamics in ballistic graphene-based short Josephson
junctions. The supercurrent through a graphene junction shows a
non-sinusoidal phase-dependence, unlike a conventional junction
ruled by the well-known d.c. Josephson relation. A
superconductor-graphene-superconductor system exhibits
superconductive quantum metastable states similar to
those present in normal current-biased JJs. We
explore the effects of thermal and correlated fluctuations on the
escape time from these metastable states, when the system is
stimulated by an oscillating bias current. As a
first step, the analysis is carried out in the presence of an
external Gaussian white noise source, which mimics the random
fluctuations of the bias current. Varying the noise intensity, it is
possible to analyze the behavior of the escape time from a
superconductive metastable state in different temperature regimes.
Noise induced phenomena, such as resonant activation and noise
induced stability, are observed. The study is extended to the case
of a coloured Gaussian noise source, analyzing how the escape time
from the metastable state is affected by correlated random
fluctuations for different values of the noise correlation time.
\end{abstract}




\maketitle

\section{Introduction}
\label{Intro}\vskip-0.2cm

\begin{figure}[b]
\vspace{-0.3cm} \centering
\includegraphics[width=87mm]{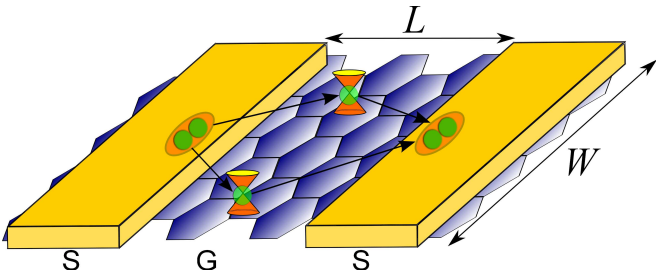}
\caption{(Color online) Schematic view of a suspended SGS device.
The electrons forming a Cooper pair, when they enter graphene, move
into different K-valleys, represented as orange cones. In the
short-junction regime, $L\ll W$.} \label{SGSDevice}
\end{figure}
The possibility of obtaining \emph{graphene}~\cite{Nov04},
by extraction of single layers from graphite, paved
the way for a new generation of superconductive
graphene-based devices. In particular, the evidence of
proximity-induced superconductivity~\cite{Hee07, Du08}, due to the
one-atom thick nature of graphene, promoted the realization of
superconductor-graphene-superconductor (SGS) structures. The
refractoriness of graphene to the surface oxidation
in natural environment favours the realization of
highly transparent contacts with the superconductive electrodes.
Furthermore, superconductivity in graphene, pure or doped, was
predicted and explored~\cite{Uch07, Fed14} and new devices, as
dc-SQUIDs~\cite{Gir09, Gir09-2}, proximity Josephson sensors~\cite{Vou10}, or
bolometers based on superconductive tunnel junction
contacts~\cite{Du14}, were fabricated using graphene. The charge
carriers in graphene are massless quasiparticle, the Dirac fermions,
with pseudo-spin half and linear energy
dispersion~\cite{Cas09}. The band structure shows contact points,
called Dirac points, beetwen the conduction and valence
bands~\cite{Cas09}. These peculiar electronic
properties~\cite{Cas09} give rise to interesting
phenomena, such as specular Andreev reflection~\cite{Bee06}, unusual
propagating modes along graphene channels~\cite{Tit07}, oscillatory
dependence of the Josephson current on the barrier thickness and
applied bias voltage~\cite{Mai07}. Titov and Beenakker~\cite{Tit06}
predicted, in the limit of zero temperature, the
behavior of critical current and current-phase relationship
(C$\Phi$R) for a short \emph{ballistic} SGS system.
Taking a cue from these results, Lambert \emph{et
al.}~\cite{Lam11} showed the existence of a plasma
frequency and a washboard potential for a suspended graphene
junction.\\
A Josephson junction (JJ) is a mesoscopic system in which
macroscopic quantities, as current and voltage, are directly
dependent on the transient dynamics of a microscopic order
parameter. The output of this device is strongly affected by
environmental perturbations, that is stochastic fluctuations of
temperature, current or magnetic field. Different aspects of
graphene-based junctions in noisy environment were already examined
by several authors. Miao \emph{et al.}~\cite{Mia09} took into
account the noise induced premature switching in underdamped SGS JJ
at finite temperature. Specifically, in
Ref.~[\onlinecite{Mia09}] the reduction of the critical current and
variations in the product $I_CR_N$ were experimentally observed and
theoretically explained considering non-negligible
the thermal fluctuations. Other
authors~\cite{Du08,Jeo11} suggested a supercurrent reduction by
premature switching induced by thermal and electromagnetic noise. 
Coskun \emph{et al.}~\cite{Cos12} systematic studied the thermally activated dynamics of phase slip in SGS JJs throught the measurement of the switching current distribution. They found an anomalous temperature dependence of the switching current dispersion due to nontrivial structure~\cite{Tit06,Hag10} of the Josephson current.
A simple stochastic model to explore the electrodynamics of an
underdamped graphene JJ was proposed by Mizuno \emph{et
al.}~\cite{Miz13}. They stressed the importance of
realizing high quality suspended SGS structures, to prevent
disorders due to the conventionally used substrates, whereby a flow
of supercurrent at high critical temperature can be obtained. The SGS junction is a good candidate for the
fabrication of gate-tunable phase qubits~\cite{Lee11,Cho13}. In
Ref.~[\onlinecite{Lee11}] the study of the stochastic switching current
distribution in a SGS junction for low temperatures allowed to
highlight the macroscopic quantum tunneling and energy level
quantization, similarly to conventional JJs. Moreover,
Lee \emph{et al.}~\cite{Lee11} studied the switching current
distribution in both quantum and thermal regime, building up a
computational analysis based on the pure resistively and
capacitively shunted junction (RCSJ) model for a conventional
JJ~\cite{Bar82}. Considering a range of temperatures in which the
dynamics is exclusively ruled by thermal fluctuations, Lee \emph{et
al.}~\cite{Lee11} observed disagreement beetwen the experimental and
fitted temperatures. To understand this discrepancy, they invoked
the misuse in the model of the pure sinusoidal Josephson current
distribution, neglecting however any noise induced
effects on the escape rate from the superconductive state.\\
Our work fits well into this assorted scenario,
since it aims to study how thermal fluctuations
affect the behavior of a SGS junction. In
particular, we study the influence of Gaussian (white or colored)
noise sources on the switching dynamics from the superconductive
metastable state to the resistive one in a
suspended graphene-based short JJ, considering the proper
C$\Phi$R~\cite{Tit06}. We recall that the effects of thermal
fluctuations on the dynamics of conventional
short~\cite{Pan04,Gor06,Gor08,Aug08,GorPanSpa08} and
long~\cite{Fed07,Fed08,Fed09,Aug09,Val14} JJs have been thoroughly
investigated, both theoretically
predicting~\cite{Pan04,Gor06,Gor08,Aug08,GorPanSpa08,Aug10,Val14}
and experimentally observing~\cite{Yu03,Sun07,Pan09} noise induced
effects in the superconductive lifetime of a JJ. The rate of switching from the JJ metastable superconducting state encodes information about the noise present in an input signal~\cite{Gra08,Urb09,Ank07,Suk07,Kop13}. The characterization of JJs as detectors, based on the statistics of the escape times, has been recently proposed~\cite{Gra08,Urb09,Fil10,Add12,Ank07,Suk07,Kop13}. Specifically, the statical analysis of the switching from the metastable 
superconducting state to the resistive running state of the JJ has been proposed to detect weak periodic signals embedded in a 
noisy environment~\cite{Fil10,Add12}.
In this paper we
explore therefore the transient dynamics of an underdamped SGS
junction, considering the simultaneous action of an external driving
force oscillating with frequency $\omega$, and a
stochastic signal which represents a random force of intensity
$\gamma$. We focus our analysis on the mean
permanence time in the superconductive state. The
study is performed fixing the initial values of the
applied bias current $i_0$ and the correlation time
$\tau_c$ of the colored noise source, and varying
the frequency $\omega$ and the noise intensity
$\gamma$. Whenever possible, we compare our results with results
obatained for normal JJs. The similarities in the
behavior of the graphene-based and normal junctions allow to
interprete our results referring to conventional JJ
quantities, such as plasma frequency (Eq.~\ref{ConvPlasmaFreq}).\\
The paper is arranged as follows. The next section includes an
overview about the physical model used. In Sec. III the statistical
features of Gaussian and correlated noise sources are examined. Sec.
IV contains the computational details. In Sec. V the theoretical
results are shown and analyzed. The Sec.VI contains a \emph{probability density function} (PDF) analysis of the escape times carried out setting the system parameters associated with the appearance of noise induced non monotonic effects in the mean switching times. In Sec. VII conclusions are drawn.

\section{The Model}
\label{Model}\vskip-0.2cm The electrodynamics of a JJ can be
explored looking at the time evolution of the order
parameter $\varphi$, that is the phase difference between the wave
functions of the two coupled superconductors forming the device.
According to the RCSJ model and including the environmental
influence, the equation of motion for $\varphi$ is
\begin{equation}
\varphi_{tt}(t)+\bJ \varphi_{t}(t)=i_b(t)-i_{\varphi}(t)+i_{f}(t)
\label{RCSJ}
\end{equation}
in which $i_b(t)$ and $i_{\varphi}(t)$ are the bias and supercurrent
respectively, both normalized to the critical current of the
junction $i_c$. The term $i_f(t)$ represents the stochastic noise
contribution. The subscripts of $\varphi$ denote partial derivatives
in time. The use of normalized variables
allows to extend, in a direct and simple way, the
theoretical results to different experimental settings.
Eq.~(\ref{RCSJ}) is in accordance with the Johnson
approach~\cite{Bar82}, since it includes a damping
parameter $\bJ=(\oP R_NC)^{-1}$, multiplied by $\varphi_{t}(t)$, and
assumes the time variable normalized to the inverse of the
zero-bias plasma frequency $\oP=\sqrt{2 \pi i_c/(\Phi_0 C)}$ ($R_N$
and $C$ are the normal resistance and capacitance of the junction,
and $\Phi_0=h/2e$ is the magnetic flux quantum).
Introducing the parameter $\beta_{_{C}}=\bJ^{-2}$,
Eq.~(\ref{RCSJ}) can be alternatively arranged in the
Stewart-McCumber framework~\cite{Bar82}, according which
a term $\beta_{_{C}}\varphi_{tt}(t)$ is included in
the equation, and the time variable is normalized to the inverse of
the JJ characteristic frequency $\omega_{c}=\omega^2_{p_{0}}R_NC$.
The JJ behavior can be depicted as the motion of a \lq\lq{}phase
particle\rq\rq{} with mass $m=C(\Phi_0/2\pi)^2$ rolling down
along the profile of a potential, called the
\emph{washboard potential}, composed by a tilted sequence of wells.
For a conventional current biased junction, the normalized
supercurrent and washboard potential have the well-known expressions
\begin{figure*}[htbp!!]
\centering
\includegraphics[width=183mm]{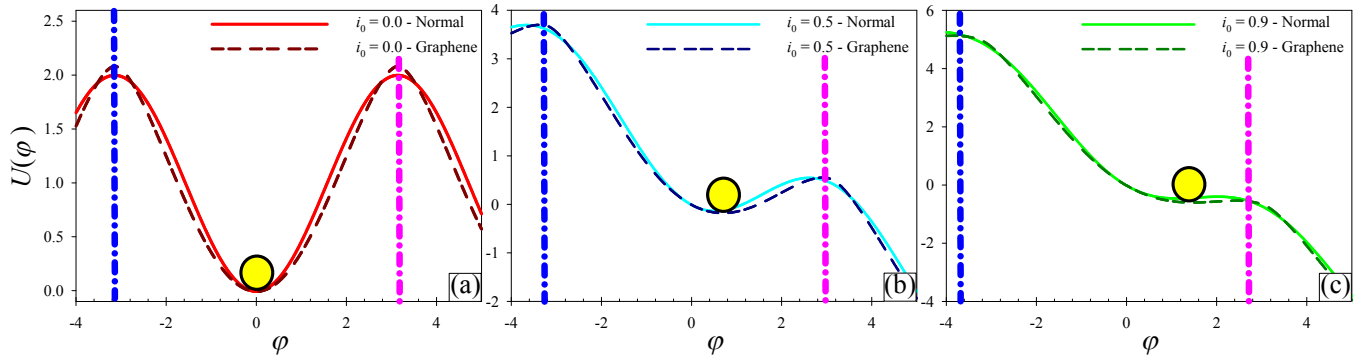}
\caption{(Color online) Washboard potential for conventional~(see
Eq.~(\ref{NormalWB})) and graphene~(see Eq.~(\ref{GrapheneWB})) JJs
(solid and dashed lines, respectively), for different initial values
of the bias current: (a) $i_0=0.0$; (b) $i_0=0.5$; (c) $i_0=0.9$. It
is also shown the initial position (bottom of the potential well) of
the \lq\lq{}phase particle\rq\rq{}. Blue and pink dotted-dashed
lines indicate the left and right absorbing barriers, respectively.}
\label{figWB}
\end{figure*}
\begin{eqnarray}\label{NormalIc}
i_{\varphi}(t)&=&\sin(\varphi(t))\\
\nonumber\\
U ( \varphi ,t )&=&-E_{J0} \left [\cos(\varphi (t))+i_b(t) \thinspace \varphi (t)\right ],
\label{NormalWB}
\end{eqnarray}
where $E_{J0}=\Phi_0 i_c/2\pi$ is the Josephson coupling energy,
that is the energy initially stored in the junction. The bias
current represents the slope of this potential. Eq.~(\ref{NormalIc})
is the \emph{d.c. Josephson relation}. In the limit of small
amplitude oscillations, the JJ plasma frequency corresponds to the
oscillation frequency in the bottom of a potential well, modified
by the presence of a bias current according to
\begin{equation}
\omega_{_{P}}(t)=\oP\sqrt[4]{1-i_b^2(t)}.
\label{ConvPlasmaFreq}
\end{equation}
Titov and Beenakker~\cite{Tit06} calculated the C$\Phi$R and
critical current for a ballistic graphene-based junction at the
Dirac point. They address the problem in the framework of the Dirac-Bogoliubov-de Gennes (DBdG) equation~\cite{Bee06, DeG66}. Considering the Josephson current at zero temperature~\cite{Bro97}
\begin{equation}
I(\varphi)=-\frac{4e}{\hbar}\frac{\mathrm{d} }{\mathrm{d} \varphi}\int_{0}^{\infty }d\varepsilon \sum_{n=0}^{\infty }\rho _n(\varepsilon ,\varphi)\varepsilon, 
\label{DBdG_Iphi}
\end{equation}
supposing an ``ideal'' normal-metal-superconductor interface and taking “infinite mass” boundary conditions~\cite{Two06} they obtained the following expressions
\begin{eqnarray}
 i_{\varphi}(t)&=&\frac{i(\varphi)}{i_c}=\frac{2}{1.33}\cos \left ( \frac{\varphi }{2} \right )\tanh^{-1}\left [ \sin \left (\frac{\varphi }{2} \right) \right ]
 \label{GrapheneIphi}\\
 i_c&=&1.33 \frac{e\Delta_0}{\hbar}\frac{W}{\pi L},
\label{GrapheneIc}
\end{eqnarray}

where $W$ and $L$ are the linear dimensions of the device~(see
Fig.~\ref{SGSDevice}), that is the length of the superconductive
plates and their separation, respectively. Furthermore $\Delta_0$ is
the superconductive excitation gap, $e$ the electron charge and
$\hbar$ the reduced Plank\rq{}s constant. The Eqs.~(\ref{GrapheneIphi}) and~(\ref{GrapheneIc}) refer to the \emph{short-junction} regime, in
which $L$ is smaller than the superconducting coherence length $\xi$, that is the distance to which a Cooper pair spreads, and for
short and wide normal region, i.e. $L\ll W$. We
recall that the simple C$\Phi$R given in Eq.~\ref{GrapheneIphi} is
obtained in the limit of zero temperature. Hagim\'asy \emph{et
al.}~\cite{Hag10} calculated a more general formula for finite
temperature and arbitrary junction length. However, an analytic
expression for the Josephson current, such as that
given in terms of washboard potential, can not be
obtained except for $T=0$. Indeed,
for vanishing temperature the expression by
Hagim\'asy \emph{et al.} correctly converges to
that obtained by Titov and Beenakker. Instead, for $T \to
T_c$, the non-sinusoidal
supercurrent derived by Hagim\'asy \emph{et al.} in both long and short junction regime, converges to a sinusoidal behavior. In the short junction limit, cf. Fig.~1a and Fig.~3a in
Ref.~[\onlinecite{Hag10}], as long as $T\lesssim T_c/4$, the
critical current and $i(\varphi)$ hardly change, so
that Titov and Beenakker's formula remains
valid~\cite{Hag14PrivateComm}. This temperature threshold can be
also deduced from the gap equation of the BCS theory, cf.~Eq.~(8) of
Ref.~[\onlinecite{Hag10}]. The work presented in
this paper is therefore strictly valid in a wide range of
temperature values, and represents a good
approximation for temperatures far from the critical value. For completeness, in the long junction limit, $L\gg W$, the Josephson current reduces to
\begin{equation}
I_{\varphi}=\frac{e\Delta }{\hbar}\tanh\left ( \frac{\Delta }{2T} \right )e^{-\pi L/W}\sin \varphi=I_c(T) \sin \varphi
\label{Iphi_LJJlimit}
\end{equation}
showing the same $\varphi$-dependence of conventional JJs (see supplemental material of Ref.~\onlinecite{Cos12}).
Lambert \emph{et al.}~\cite{Lam11} proposed, for the Titov's
Josephson current, the following washboard-like
potential
\begin{eqnarray} \label{GrapheneWB}
\nonumber \tilde{U}(\varphi,t ) = &-&E_{J0}  \Biggr\{ -\frac{2}{1.33} \biggr\{ 2\sin\left (\frac{\varphi }{2}  \right )\tanh^{-1}\left [ \sin\left (\frac{\varphi }{2} \right ) \right ] + \\
&+&\ln \left [ 1- \sin^2\left (\frac{\varphi }{2}\right )\right]  \biggr\} +i_b(t)\varphi \Biggr\}.
\end{eqnarray}
The analytic knowledge of the potential allows to well impose
the initial condition and the thresholds for the escape time
calculations. As well as the conventional $U(\varphi,t )$ (see
Eq.~\ref{NormalWB}), the potential $\tilde{U}(\varphi,t )$
consists of a tilted sequence of wells. The
position of the phase particle or, more precisely, its dynamical
condition along this potential, defines the working regime of the
junction. In the \emph{superconductive state} the particle lies in a
well, while in the \emph{resistive state} it rolls down along the
potential. When this happens, a non-zero mean voltage $V$ across the
junction appears, according to the \emph{a.c. Josephson relation},
$\varphi_t=2\pi V/\Phi_0$. Furthermore, depending
on the damping parameter value, the \emph{phase
diffusion state}, that is an escape event with a retrapping in the
first subsequent minimum, could be established.
When $i_b(t)\geq1$, that is when the applied bias current exceeds
the critical value, both potentials (Eqs.~(\ref{NormalWB})
and~(\ref{GrapheneWB})) lose their \lq\lq{}maxima and minima\rq\rq{}
structures and the particle tends to freely slip.\\
We explore the response of the system to the simultaneous action of
both d.c. and a.c. current
sources. The bias current, composed by a constant term, $i_0$,
representing its initial value, and an oscillating part whose
frequency $\omega$ is normalized to $\oP$, is
therefore given by
\begin{equation}
 i_{b}(t)=i_0+A\sin(\omega t).
\label{BiasCurrent}
\end{equation}
\begin{table}
\vskip0.1cm
\begin{center}
\begin{tabular}{lc|c|c|c|c|c|c|}
&&Mizu-&Coskun~\cite{Cos12}&Du~\cite{Du08}&Heer-&English~\cite{Eng13}&Miao~\cite{Mia09}\\&&no~\cite{Miz13}&&&schee~\cite{Hee07}&\scriptsize{Samples}&\\&&&&&&\scriptsize{A/B/C/D}&\\ \hline
$i_{C}$&mA&100&10&800&10&71/107&110\\&&&&&&39/160&\\ \hline
$\beta_{C}$&&76&16&&&&\\ \hline
C&pF&1&12-50&&&&\\ \hline
$R_{N}$&$\Omega$&500&10&&&&\\ \hline
$T$&K&3&0.4&0.2&0.3&0.01&0.3\\ \hline
$T_{CO}$&K&0.02&[8-17]$\cdot10^{-3}$&&&&0.12-1.2\\ \hline
$\gamma_{C}$&&1.3&1.7&0.01&1.3&6/4/11/3&0.11\\&&&&&&$\cdot10^{-3}$&\\  \hline
$\omega_{P0}$&GHz&17&0.8-1.6&&&&$10^2$-$10^3$\\ \hline
\end{tabular}
\end{center}
\caption{Experimental values of different JJ parameters, calculated or directly acquired by various published works.}\label{ExpValuesTable}
\end{table}
By choosing properly the values of $i_0$ and $A$,
within a period it is possible to achieve values of
$i_b(t)$ greater than $1$. A direct comparison
between the potentials for normal and graphene-based
JJ is given in Fig.~\ref{figWB} for $i_0=$ 0.0 (panel a), 0.5
(panel b), 0.9 (panel c). Here it is worth noting that differences,
though small, between the graphene and
normal JJ curves are detectable. Fig.~\ref{figWB}
shows also the initial condition for the fictitious particle, which
is located in the potential minimum. We can assume that the system
leaves the superconductive regime when the particle reaches one of
the nearest maxima. Two absorbing barriers are therefore placed in
correspondence to these maxima, as highlighted in Fig.~\ref{figWB}
(see dotted-dashed lines). Recording for each
realization the escape times $t^{esc}$, i.e. the time required to
pass a barrier, for an enough large number $N$ of
realizations, the \emph{mean first passage time}
(MFPT) is defined as
\begin{equation}
\tau=\frac{1 }{N}\sum_{i=1}^{N}t^{esc}_i.
\label{MFPT}
\end{equation}
The oscillating force acting on the system,
$i_b(t)$, and stochastic fluctuations, $i_f(t)$, due to the
environmental influence, drive the switching
dynamics. Two different mechanisms can therefore
cause overcome of the potential barrier: the macroscopic quantum
tunneling or the thermally activated passage. These processes are
triggered in distinct ranges of temperature so that,
for vanishing values of the bias and damping, a threshold value
exists, $T_{CO}=\hbar \oP / 2\pi k$ ($k$ is the Boltzmann constant),
called \emph{crossover temperature}. In a damped system,
when a polarization current is applied, this value
is slightly reduced, becoming~\cite{Gra84}:
\begin{equation}
\TsCO=\hbar \omega_R / 2\pi k, \label{Tcrossover}
\end{equation}
where $\omega_R=\omega_P\left \{ \sqrt{1+\alpha ^2}-\alpha  \right
\}$, $\alpha=(2\omega_PR_NC)^{-1}\propto\bJ$. For $T<\TsCO$ the
system undergoes a \emph{quantum tunneling regime}. On the other
hand, for $T>\TsCO$, the system works in the \emph{thermal
activation regime}. Here we do not take into account quantum
effects. In this condition, when thermal fluctuations are neglected,
the phase can remarkably change merely as the applied current
approaches the critical value $i_c$ (the system moves into a
resistive regime). Conversely, considering noise effects,
transitions along the potential can occur also applying a current
much smaller than $i_c$. As already pointed out, the phase dynamics
is affected by dissipative phenomena, responsible
for peculiarities of the system, ranging from overdamped (high
viscosity $\beta_{_{J}}\gg 1$) to underdamped (low viscosity
$\beta_{_{J}}\ll 1$) condition.
The Table~\ref{ExpValuesTable} shows a collection of few experimental values, for different graphene-based JJs, calculated or, whenever possible, directly acquired by different published works~\cite{Cos12,Du08,Eng13,Mia09,Miz13,Hee07}. Blank cells indicate that the value of the related variables are not available. The values of the parameters $\beta_{_{C}}=\bJ^{-2}$ suggest that these systems often~\cite{Cos12,Miz13} work in underdamped conditions. Moreover, the comparison between the working temperature $T$ and the crossover value $\TsCO$, underlines the thermally activated behavior of the switching dynamics characterizing these junctions~\cite{Cos12,Miz13,Mia09}.\\

\emph{The noise source}. $-$ An exhaustive analysis of a real
device has to take into account environmental fluctuations
continuously affecting the system, such as
unpredictable variations of current and temperature. Thus the
deterministic RCSJ model can be improved by
considering the presence of the stochastic current $i_f$~(see
Eq.~\ref{RCSJ}), in a first approximation modeled
using a Gaussian \lq\lq{}white\rq\rq{} noise source.
The stochastic non-normalized current $I_f\left (
\tilde{t}\right )$ is therefore characterized by the well-known
statistical properties of a Gaussian random process
\begin{eqnarray}
\left \langle I_f\left (  \tilde{t}\right ) \right \rangle = 0
\hspace{6 mm} \left \langle I_f\left (  \tilde{t}\right )I_f\left (
\tilde{t}+\tilde{\tau}\right ) \right \rangle =
2\frac{kT}{R_N}\delta \left ( \tilde{\tau} \right ),
\label{WNProperties}
\end{eqnarray}
where $T$ is the temperature. Using normalized
current and time, the correlation function becomes
\begin{equation}
\left \langle i_f(t)i_f(t+\tau ) \right \rangle = 2\gamma (T)\delta
\left ( \tau  \right ), \label{WNCorrelation}
\end{equation}
where the dimensionless amplitude $\gamma (T)$ is proportional to
the temperature $T$. We note that the expression of
$\gamma (T)$ depends on the approach used to manage
Eq.~(\ref{RCSJ})
\begin{subequations}\label{WNAmp}
\begin{align}
\mathrm{McCumber})\hspace{5mm}\gamma^c (T)&=\frac{kT}{R_N}\frac{\omega_{c}}{i^2_c}=\frac{2e}{\hbar}\frac{kT}{i_c}=\frac{kT}{E_J}\label{WNAmpA}\\
\mathrm{Johnson})\hspace{5mm}\gamma^p (T)&=\frac{\oP}{\omega_{c}}\gamma^c (T)\label{WNAmpB}
\end{align}
\end{subequations}
It is worth noting that the noise intensity can be
also expressed as the ratio between the thermal and Josephson
coupling energies~(see Eq.~(\ref{WNAmpA})). Few $\gamma^c $ values, calculated for several experimental setting, are shown in the Table~\ref{ExpValuesTable}.
More in general, $i_f(t)$ can represent a Gaussian
colored noise, modeled as an exponentially correlated noise source.
Specifically, in this work the noise source is
described by the well-known Ornstein-Uhlenbeck (OU)
process~\cite{Gar04}
\begin{equation}
d i_f(t)=-\frac{1}{\tau _c}i_f(t)dt+\frac{\sqrt{\gamma }}{\tau
_c}dW(t), \label{OUprocess}
\end{equation}
where $\gamma$ and $\tau _c$ are the intensity and correlation time
of the noise source, respectively, and $W(t)$ is the Wiener process,
characterized by the well-known statistical properties: $\left
\langle dW(t) \right \rangle=0$ and $\left \langle dW(t)dW(t')
\right \rangle=\delta \left ( t-t' \right )dt$.

The correlation function of the OU process is
\begin{equation}
\left \langle i_f(t)i_f(t') \right \rangle=\frac{\gamma }{2\tau
_c}e^{-\frac{\left | t-t' \right |}{\tau _c}}, \label{OUcorrfunct}
\end{equation}
and gives $\gamma \thinspace \delta(t-t')$ in the
limit $\tau_c \rightarrow 0$.\\

\begin{figure*}[htbp!!]
\centering
\includegraphics[width=180mm]{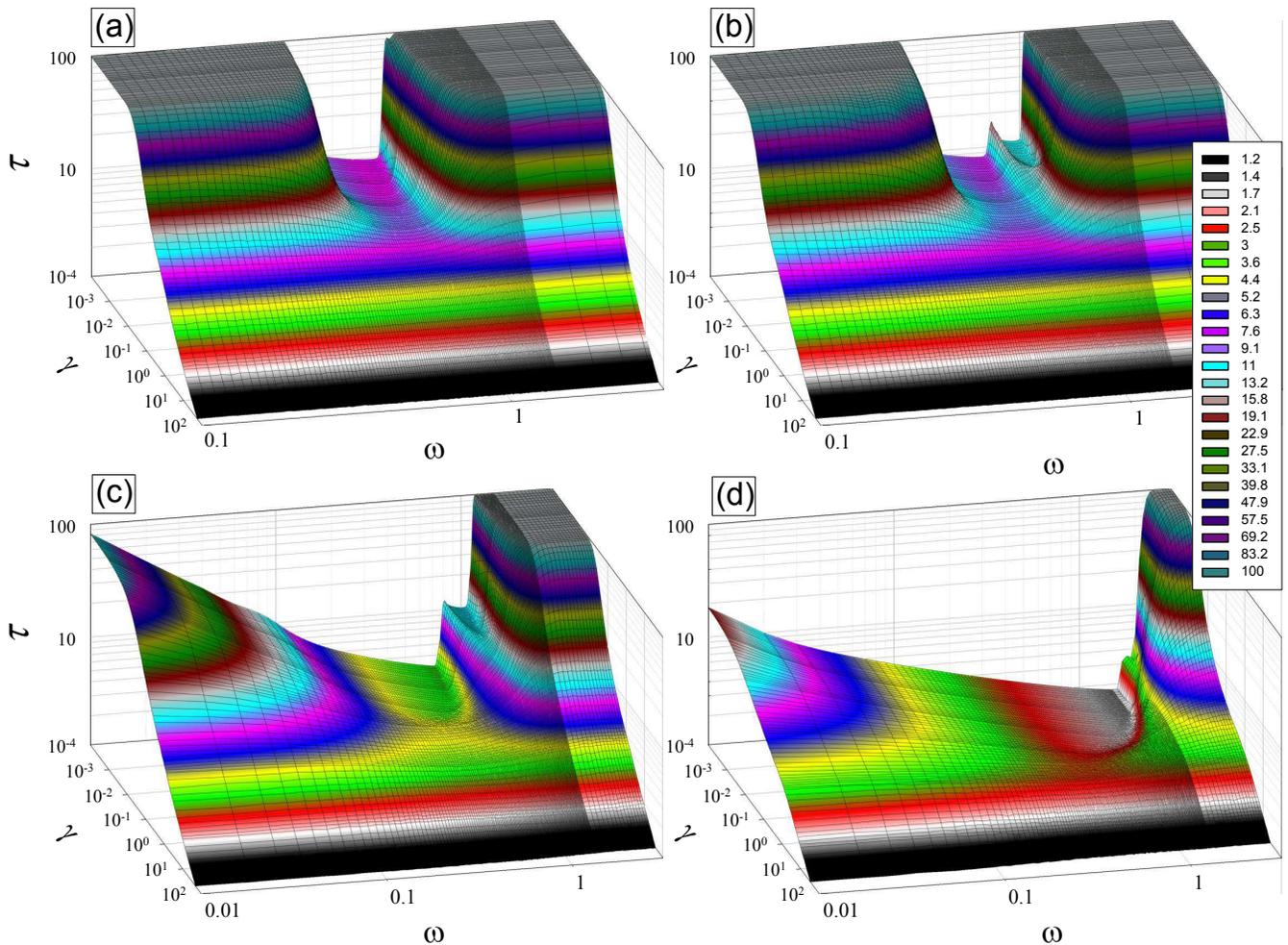}
\caption{(Color online) MFPT as a function of both $\omega$ and
$\gamma$, for $\tau_c=0.0$ and different initial values of the bias
current: (a) $i_0=0.0$ (no slope); (b) $i_0=0.1$ (small slope); (c)
 $i_0=0.5$ (intermediate slope); (d) $i_0=0.9$ (high slope).
 The legend in panel d refers to all pictures.} \label{GrapheneData}
\end{figure*}
\emph{Computational details}. $-$ The stochastic dynamics of the system is explored integrating
Eqs.~(\ref{RCSJ}) and~(\ref{OUprocess}) by a finite difference
method. Specifically, the stochastic differential
equation~(\ref{OUprocess}) is integrated within the Ito scheme. The
time step is fixed at $\Delta t=10^{-3}$ and the maximum time,
for which equations are integrated, is
$t_{max}=100$, i.e. a time large enough to catch every nonmonotonic
behavior. A collection of first passage times is obtained iterating
the procedure for a sufficiently large number of
realizations $N=10^4$. The initial condition to
solve Eq.~(\ref{RCSJ}) is set at the bottom of a valley of the
potential given in Eq.~(\ref{GrapheneWB}), closer to $\varphi=0$.
During the oscillation of the potential the two
absorbing barriers change their position, following the
displacements of the neighboring maxima. The analysis is performed
in the underdamped regime, setting $\bJ=0.1$ (corresponding to
$\beta_{_{C}}=100$). Four different values of $i_0$,
in the range $0\leq i_0< 1$, are used. The time periodical component
of $i_b(t)$, oscillates with values of the frequency $\omega$
ranging within the interval $[0.01-10]$. In our analysis the
intensity $\gamma$ of the colored noise source $i_f(t)$ varies in
the range $[10^{-4}-10^2]$, with the correlation time, $\tau_c$, set
at different values.

\section{The Analysis}

The analysis is performed studying the behavior of
the MFPT, $\tau$, as a function of the noise
intensity $\gamma$ and frequency $\omega$ of the oscillating term
in the bias current. In Eq.~(\ref{BiasCurrent}) $i_0
= 0.0, 0.1, 0.5, 0.9$, corresponding to vanishing, small,
intermediate and high values, respectively, of the initial slope of
the washboard potential. The slope of the
potential, that is the value of $i_b(t)$, is directly
related to the height of the potential barriers, so that, increasing
the value of $i_b(t)$, the right barrier's height decreases, getting
zero when $i_b(t)\ge1$. The normalized amplitude of the oscillating
term of the bias current is set at $A=0.7$ in all
numerical realizations.

We show the values of $\tau$ in
three-dimensional plots to highlight the
simultaneous presence of different nonmonotonic effects. The values
of $\gamma$ are proportional, through Eqs.~(\ref{WNAmp}), to the
temperature of the system, so that varying the noise
intensity in the interval $\gamma=[10^{-4} -10^2]$
corresponds to explore a wide range of
temperatures. The noise amplitude values calculated in different frameworks and presented in the Table~\ref{ExpValuesTable}, fall within this range. The values of the frequency $\omega$
are chosen in such a way to investigate different regimes of
alternate current: i) quasi-direct current ($\omega\ll1$); ii)
high-frequency alternate current ($\omega\gg1$); iii) alternate
current oscillating at the characteristic plasma frequency of a
conventional junction ($\omega=1$). Recalling that the driving frequency is normalized to the plasma frequency, the values of $\omega_{P0}$ included in Table~\ref{ExpValuesTable} makes it possible to give a quantitative estimation to the values taken by $\omega$. The correlation time of the
colored noise source takes the values $\tau_c= 0$
(i.e. white noise), $1, 5, 10$. The results, shown
in Fig.~\ref{GrapheneData}, were obtained using a white noise
source, that is setting $\tau_c= 0.0$, and for different values of
the initial bias current (slope of the potential), i.e. $i_0$ =
$0.0$ (panel a), $0.1$ (panel b), $0.5$ (panel c), $0.9$ (panel d).
First we can note that an overall lowering of
$\tau$ values occurs, as $i_0$ increases.
In other words, changes in the maximum slope of the
potential cause modifications in the height of the barriers (see
Fig.~\ref{figWB}). The presence of two absorbing barriers allows to
take into account the complete evolution of the phase particle from
the initial state. Considering highly tilted potential profile
(panel c of Fig.~\ref{figWB}), the particle rolls down exclusively
overcoming the right barrier. Instead, with small value of the
initial bias current (panel a of Fig.~\ref{figWB}), the possibility
of escaping over the left-side barrier causes
interesting phenomena. In particular, for $i_0=0.0$, the height of
the left and right barriers takes on the same
values within an oscillation period, so that the particle can escape
through the left or right barrier with equal probability. In all
panels of Fig.~(\ref{GrapheneData}) it is evident a
nonmonotonic behaviour, characterized by a minimum,
which indicates the presence of a \emph{resonant activation} (RA)
phenomenon~\cite{Doe92, Man00, Dub04, Man98, Pec94, Mar96, Dyb09,
Miy10, Has11, Fia11}. This effect is robust enough
to be detected in a large range of $\gamma$ values, even if it tends
to be suppressed (the minimum in the curves of MFPT \emph{vs}
$\omega$ is less pronounced) as the intensity, $\gamma$, of thermal
fluctuations increases. In particular, two different kinds of RA
can be distinguished: i) the \emph{dynamic resonant activation},
which occurs as the driving frequency approaches
the natural characteristic frequency of the system, coinciding for a
JJ with the plasma frequency~\cite{Dev84,Dev85,Mar87}; ii) the
\emph{stochastic resonant activation}, which occurs
for driving frequency close to the inverse of the average escape
time at the minimum, i.e. the mean escape time over
the potential barrier in the lowest
configuration~\cite{Add12,Pan09}. 
\begin{figure}[t]
\centering
\includegraphics[width=90mm]{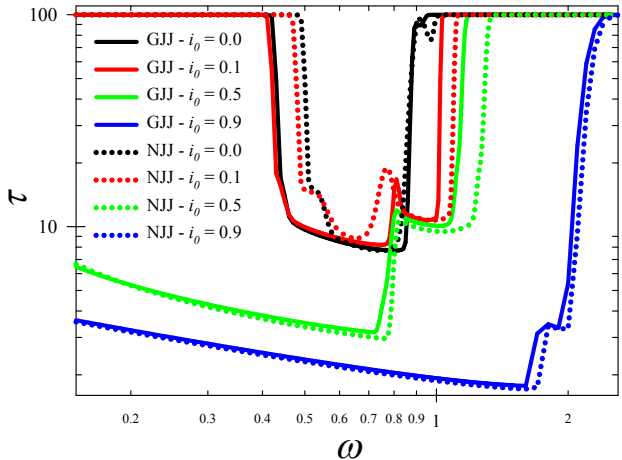}
\caption{(Color online) MFPT as a function of $\omega$, for
$\gamma=10^{-4}$, $\tau_c=0.0$, and different initial values of the
bias current: $i_0= 0.0, 0.1, 0.5, 0.9$. Solid and dotted lines
represent results for a graphene-based JJ (indicated as GJJ) and a
normal JJ (indicated as NJJ), respectively.} \label{DynamicRA}
\end{figure}
The dynamic RA is evident only in
quasi-deterministic regime, i.e. $\gamma\ll 1$, when the dynamics
depends mainly by the geometry and symmetry of the system.
Increasing the noise intensity, the stochastic RA tends to overcome
every dynamic RA effect. Fig.~\ref{DynamicRA} shows the behaviour of
the MFPT \emph{vs} $\omega$, with the noise
intensity fixed at such a value ($\gamma=10^{-4}$) that the dynamic
RA effect can be clearly observed and studied as a function of the
initial bias current $i_0$. 
\begin{figure*}[htbp!!]
\centering
\includegraphics[width=180mm]{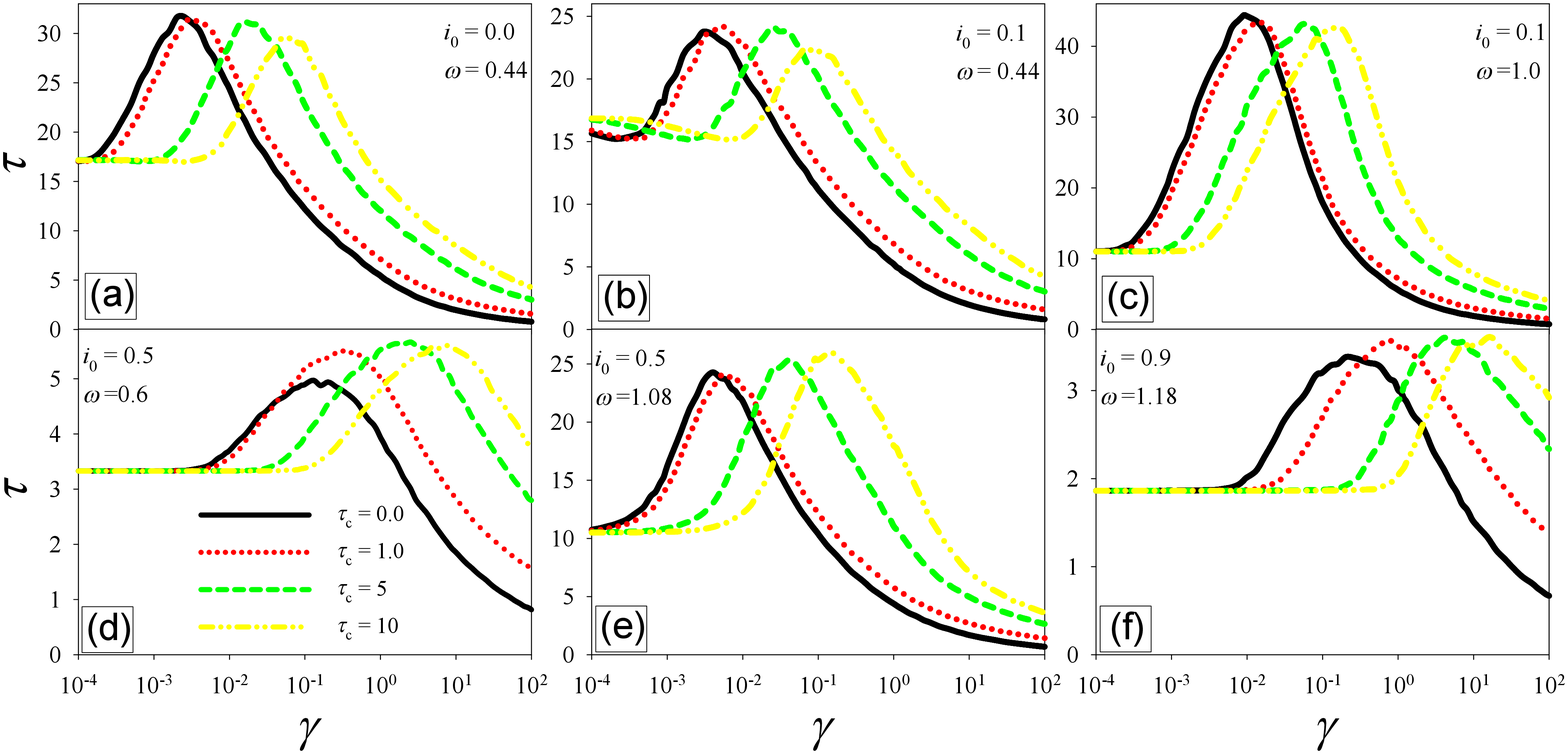}
\caption{(Color online) MFPT as a function of $\gamma$, for
different values of $\omega$, $i_0$ and $\tau_c$. In detail: (a)
$i_0=0$, $\omega=0.44$; (b) $i_0=0.1$, $\omega=0.44$; (c) $i_0=0.1$,
$\omega=1.0$; (d) $i_0=0.5$ and $\omega=0.6$; (e) $i_0=0.5$,
$\omega=1.08$; (f) $i_0=0.9$, $\omega=1.18$. The legend in panel d
refers to all pictures.} \label{NES}
\end{figure*}
More in detail,
Fig.~\ref{DynamicRA} displays results obtained for graphene-based
(solid lines) and normal (dotted lines) JJs . The MFPT values in
correspondence of the RA minima are almost junction-type
independent, even if the RA valleys for normal JJ
are shifted towards higher frequencies. In particular, the dynamic
RA minima for small $i_0$ become narrower passing
from SGS to normal junction. In
Fig.~\ref{GrapheneData}, where the noise intensity is set at
$\gamma=10^{-4}$, for $i_0=0.0$ (panel a) a single-minimum dynamic
RA is present in $\wRAa\simeq 0.81$. The RA effect becomes more
structured, slightly increasing the initial bias current.
Indeed, for $i_0=0.1$ the same effect occurs with
the presence of two minima located at $\wRAb\simeq 0.75, 0.95$ (see
panel b of Fig.~\ref{GrapheneData}). These minima are connected with
two resonance phenomena occurring in the system. Specifically, the
oscillating potential can ``tune" with the plasma oscillations for
two different values of $\omega$, one corresponding to escape events
towards left, which occur at the lowest slope, the other one
corresponding to escape events towards right, which occur at the
highest slope. This double-resonance effect can be
further explained, noting that non-vanishing values of the bias
current ($i_0\ne0$) introduce an asymmetry, e.g. with $i_0=0.1$
the highest and lowest slope are respectively
$\left | i_b(\Tp/4) \right |=0.8$ and $\left | i_b(3\Tp/4) \right
|=0.6$, where $\Tp$ is the oscillation period. For these
configurations the plasma frequencies, calculated according to
Eq.~(\ref{ConvPlasmaFreq}), are
$\omega_{_{P}}^{0.1}(\Tp/4)\simeq0.77$ and
$\omega_{_{P}}^{0.1}(3\Tp/4)\simeq0.90$. These
values, even if they do not coincide, are very close to the
frequencies $\wRAb$ for which the RA minima are observed. The small
discrepancies between $\wRAb$ and the two frequencies
$\omega_{_{P}}^{0.1}(\Tp/4)$ and $\omega_{_{P}}^{0.1}(3\Tp/4)$ can
be related to the fact that the conventional JJ
plasma frequency was used. Due to the symmetry of
the potential for $i_0=0.0$ respect to the horizontal position, the
minima observed for $i_0=0.1$ seem to merge in the
larger minimum located, for vanishing bias current,
at $\wRAa\simeq0.81$ (see Fig.~\ref{DynamicRA}). 

Indeed in this
situation, the highest and lowest slope have the
same absolute value, $\left | i_b(\Tp/4) \right |=\left |
i_b(3\Tp/4) \right |=0.7$. Accordingly, in these
configurations the plasma frequencies take on values
($\omega_{_{P}}^{0.0}(\Tp/4)=\omega_{_{P}}^{0.0}(3\Tp/4)\simeq0.85$),
very close to that for which the RA minimum is observed. The
suppression of the dynamic RA, as the noise
intensity increases, is evident in the curves
obtained for $i_0=0.1$. In particular, the stochastic RA emerges at
$\gamma_{_{sRA}}^{0.1}\simeq0.005$, with the minimum located in
$\omega_{_{sRA}}^{0.1}\simeq0.7$. In these conditions the dynamics
is exclusively ruled by the noise fluctuations which ``do not see''
the potential details. Using these small values of bias current, a
trapping phenomenon occurs for $\omega\ge1$. This effect
is due to the inability of the particle to leave a
minimum, since the frequency of the oscillating
potential is larger than the characteristic frequency of the well,
i.e. the plasma frequency. These trapping phenomena however
disappear for higher values of the noise intensity.
For $i_0=0.5, 0.9$ the potential is tilted enough to
become, in the lowest configuration, well-free. If $i_0=0.5$ the
double-minimum dynamic RA is still present around the frequencies
$\wRAc\simeq 0.72, 1.02$, but the MFPT value in the first RA minimum
is smaller than that calculated for $i_0=0.1$. 
This is due to the fact that for $i_0=0.5$ the virtual particle,
i.e. the phase difference between the wave functions
of the two superconductors, is able to leave the potential well in
a shorter time, escaping through the right potential barrier. The
slope $i_0=0.5$ in fact is sufficient to produce a
right-side escape event already after a quarter of
an oscillation period (indeed $\tau\simeq \Tp/4$), whereas for
$i_0=0.1$ the particle needs one complete oscillation to pass the
same barrier. On the other hand, the values of $\tau$ in the second
RA valley for $i_0=0.1$ and $i_0=0.5$ are almost equal, since the
particle needs more than one complete oscillation (for both slopes
$\tau\simeq \Tp+3\Tp/4$) to escape from the left potential barrier.
Setting $i_0=0.9$, the dynamic RA is just hinted and only the
minimum around $\wRAd\simeq1.6$, corresponding to a
highly sloping potential, is clearly detectable. 
Trapping phenomena
at high frequencies are still present. Specifically
they appear for frequencies larger than the following threshold
values: $\omega^{0.5}_{thr}\simeq 1.2$ and
$\omega^{0.9}_{thr}\simeq 2.4$. Increasing the value of the bias
current, the right potential barrier decreases. As a
consequence, trapping phenomena can occur only if the potential
oscillates at higher frequencies. Furthermore, the parabolic
approximation (linearization of the potential at the
bottom of the well) used to calculate the plasma frequency (see
Eq.~(\ref{ConvPlasmaFreq})) fails for a highly tilted potential. In
Fig.~\ref{GrapheneData} we show the behaviour of the MFPT as a
function of the noise intensity $\gamma$. 
In all panels of Fig.~\ref{GrapheneData} we note the presence of another
noise induced effect, known as \emph{noise enhanced stability}
(NES)~\cite{Val14,Dub04,Man96,Agu01,Spa04,DOd05,Fia05,Hur06,Spa07,Man08,Yos08,Fia09,Tra09,Fia10,Li10,Smi10}.
Indeed the curves of $\tau$ \emph{vs} $\gamma$ are
characterized by a nonmonotonic behavior with the presence of a
maximum. This nonmonotonic behavior is different from
that expected from the Kramers theory and its
extensions~\cite{Kra40,Mel91,Han90}. The enhancement of stability
present in the curves of Fig.~\ref{GrapheneData}, first noted by
Hirsch et al.~\cite{Hir82}, has been observed in different physical
and biological systems, and belongs to a highly topical
interdisciplinary research field, ranging from condensed matter
physics to molecular biology and cancer growth
dynamics~\cite{Spa07,Spa12}. More in detail, we
note that the $\tau$ \emph{vs} $\gamma$ behaviour shows the presence
of NES for any frequency taken in an interval around the different
frequencies $\omega _{_{dRA}}$. This suggests that
the origin of this nonmonotonic effect can lie in the resonance
phenomenon, involving the plasma frequency, previously discussed
about the RA effect. Specifically, for $i_0=0.0$
this effect occurs for $\omega_{_{NES}}\in[0.43-0.87]$. For each
value $i_0=0.1, 0.5$ of the bias current, there are two $\wRA$ frequencies and, correspondingly, two different ranges of frequencies giving evidence of NES effects.\\
In detail: $\omega^{(1)}_{_{NES}}\in[0.42-0.78]$ and
$\omega^{(2)}_{_{NES}}\in[0.84-1.02]$ for $i_0=0.1$, and
$\omega^{(1)}_{_{NES}}\in[0.24-0.77]$ and
$\omega^{(2)}_{_{NES}}\in[0.97-1.14]$ for $i_0=0.5$.\\
Using highly tilted potential, i.e. $i_0=0.9$, there is only one RA
minimum and, according to the correspondence previously observed,
only one range of frequencies ($\omega_{_{NES}}\in[0.4-2.4]$) for
which the NES phenomenon is found. According to this
analysis, the curves of Fig.~\ref{NES}, obtained for different
values of the noise correlation time ($\tau_c= 0.0, 1.0, 5, 10$,)
show the presence of NES for values of $\omega$ chosen in the
intervals given above. In all curves,  as $\tau_c$
increases, the maxima are shifted towards higher values of the noise
intensity. Moreover, the MFPT values around the NES maxima tend to
slightly reduce for low slopes (small values of $i_0$) of the
oscillating potential (panels a, b and c of Fig.~\ref{NES}) and to
increase for high slopes (panels d, e and f of the Fig.~\ref{NES}).
\begin{figure}[b]
\centering
\includegraphics[width=89mm]{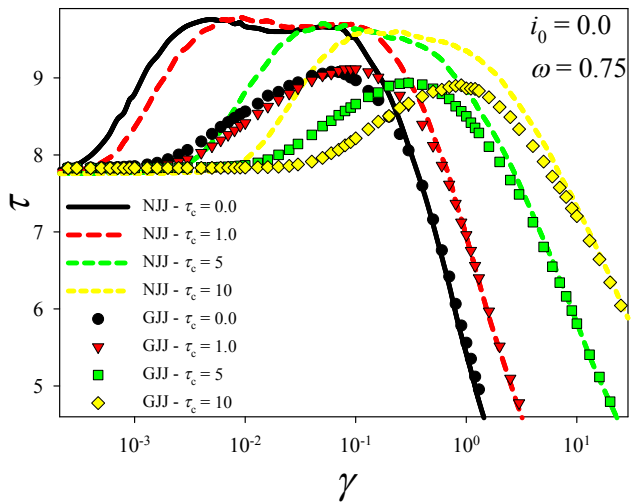}
\caption{(Color online) MFPT as a function of $\gamma$, for
$\omega=0.75$, $i_0=0.0$, and different values the noise correlation
time: $\tau_c=0.0, 1.0, 5, 10$. Lines and symbols represent results
for a normal JJ (NJJ) and a graphene-based JJ (GJJ), respectively.}
\label{NES_Comparison}
\end{figure}
These features, i.e. the shift towards higher frequencies and
modification in the maxima of MFPT for increasing values of
$\tau_c$, are present also in a
conventional JJ. In Fig.~\ref{NES_Comparison},
where $i_0=0.0$ and $\omega=0.75$, it is possible to observe that
for a normal JJ respect to a graphene junction: i) the NES maxima
are broadener; ii) the phase particle remains confined in the
potential well for longer time, i.e. the $\tau$ values are slightly
higher (in accordance with the results of
Fig.~\ref{DynamicRA}); iii) the NES effect appears for lower noise
intensities. Conversely, the behaviors of normal and graphene JJs
coincide for larger values of the noise intensity $\gamma$, since
the specific potential profile becomes irrelevant due to the
strength of random fluctuations.

\begin{figure*}
\centering
$\vcenter{\hbox{\includegraphics[width=125mm]{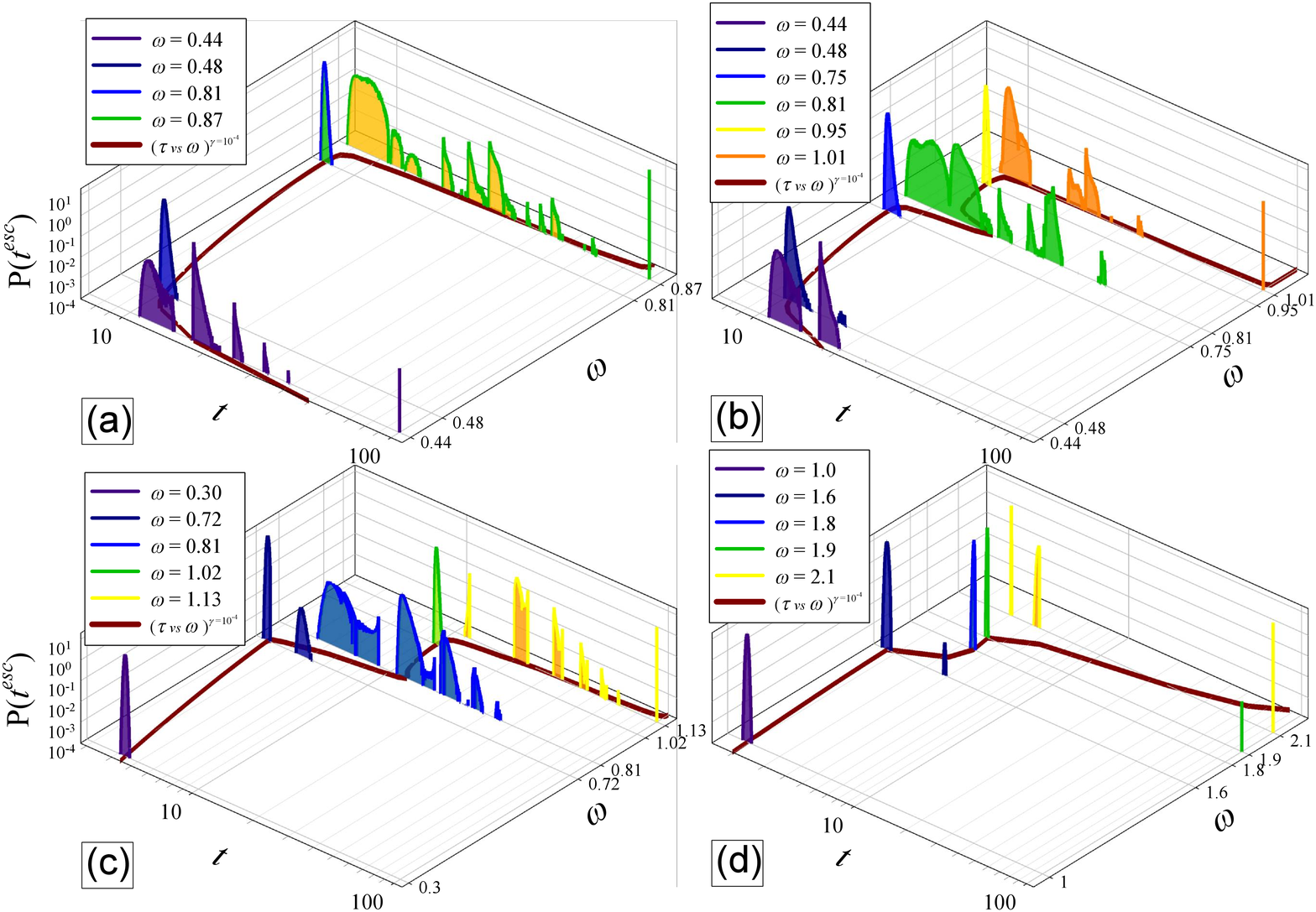}}}$
\hspace{0mm}
$\vcenter{\hbox{\includegraphics[width=52mm]{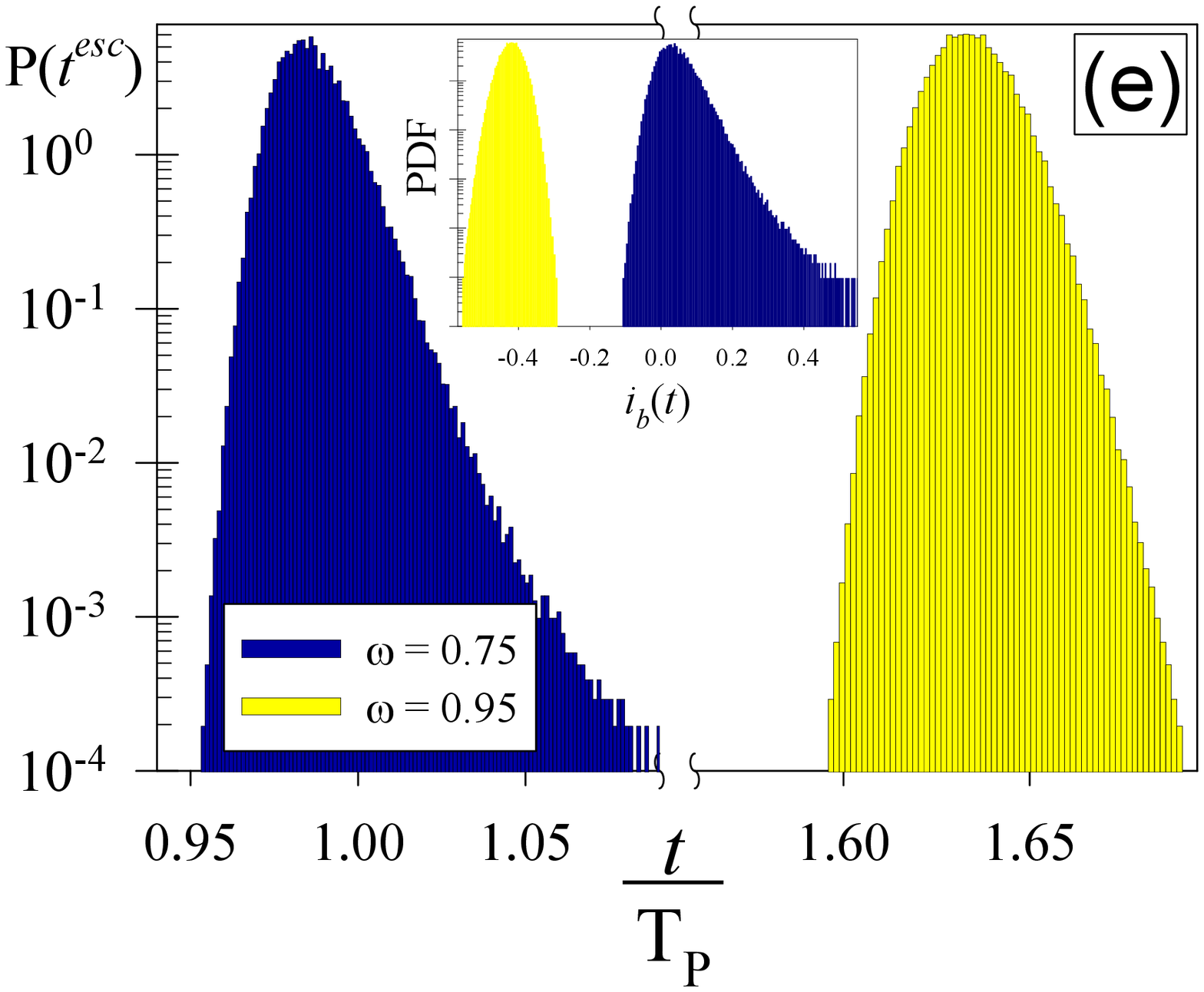}}}$
\caption{(Color online) Panels (a), (b), (c) and (d): PDFs as a function of the time $t$, varying $\omega$. Every picture is obatined fixing the values of $\gamma=10^{-4}$, $\tau_c=0$ and $i_0=\{$(a)$i_0=0$, (b)$i_0=0.1$, (c)$i_0=0.5$, (d)$i_0=0.9\}$. The MFPT \emph{versus} $\omega$ curves corresponding to the dynamic RA effects (see solid lines in Fig.~\ref{DynamicRA}) are also shown. The PDF and $t$ axes are logarithmic. Panel (e): Semi-log plot of the PDFs as a function of the time $t$, normalized to the washboard oscillation period $\Tp$, setting $i_0=0.1$ and $\omega=\wRAb=\{0.75, 0.95\}$. The inset shows the same PDF data in function of the bias current $i_b(t)$.}
\label{PDF_RA}
\end{figure*}

\begin{figure*}[t]
\centering
\includegraphics[width=180mm]{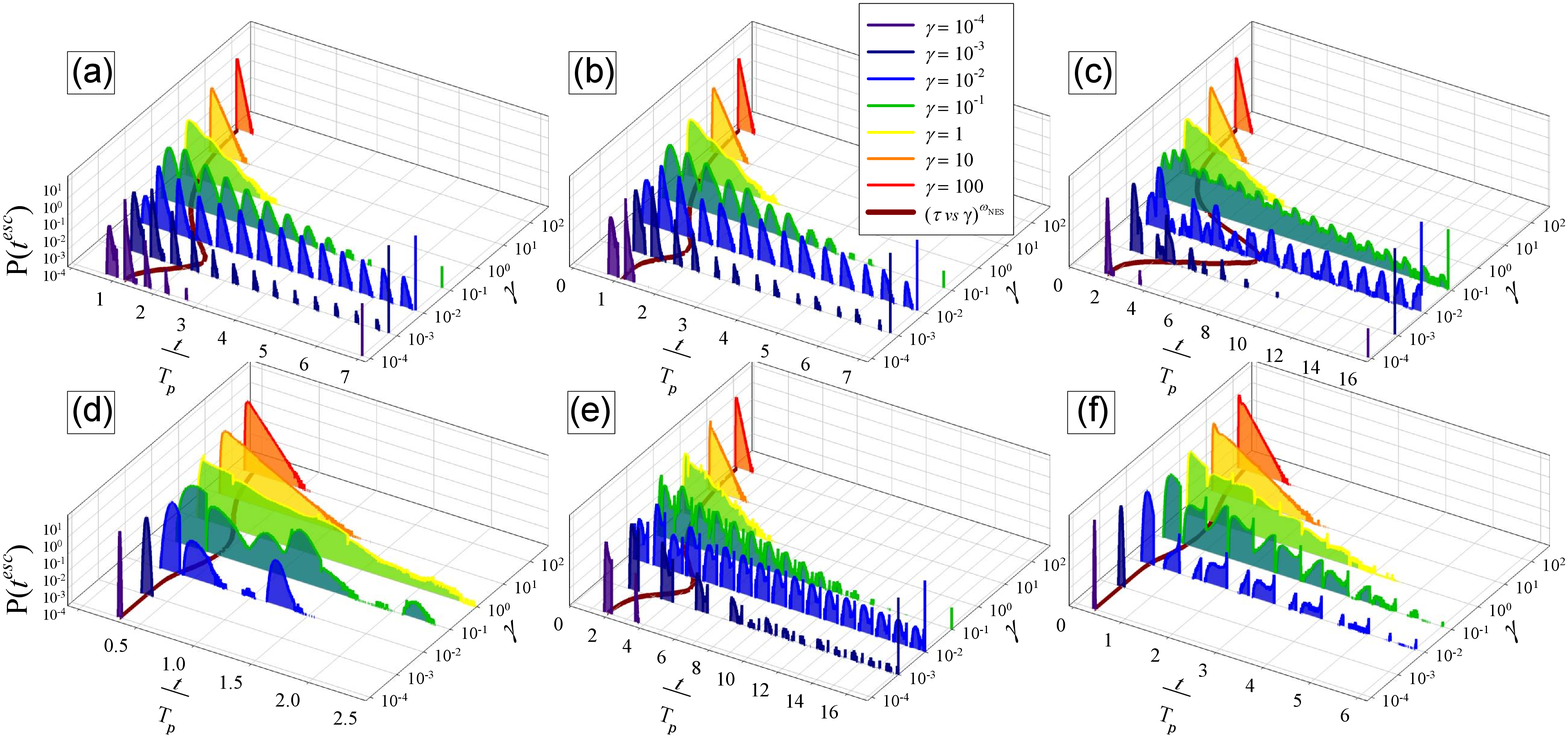}
\caption{(Color online) PDF as a function of the time $t$, normalized to the washboard oscillation period $\Tp$, varying $\gamma$. Every picture is obatined fixing the values of $\omega$, $i_0$ and $\tau_c=0$. In detail: (a)
$i_0=0$, $\omega=0.44$; (b) $i_0=0.1$, $\omega=0.44$; (c) $i_0=0.1$,
$\omega=1.0$; (d) $i_0=0.5$ and $\omega=0.6$; (e) $i_0=0.5$,
$\omega=1.08$; (f) $i_0=0.9$, $\omega=1.18$. Every picture shows also the MFPT \emph{versus} $\gamma$ curve corresponding to NES effect (see solid lines in Fig.~\ref{NES}) obtained using the same values for the other parameters. The PDF and $\gamma$ axes are logarithmic. The legend in panel (b) refers to all pictures.}
\label{PDF_NES}
\end{figure*}

\section{Probability Density Functions}
\label{PDFsection}\vskip-0.2cm
To deeply understand the behavior of the MFPT, we expand further the theoretical analysis discussing the PDFs of the switching times $\Pe$. We set the parameters of system and noise source to obtain nonmonotonic effects in the MFPT data. Every PDF is costructed implementing $N_{exp}=10^{7}$ experiments, and every curve is normalized to unity. Whenever possible, the time $t$ will be normalized to the washboard oscillation period $\Tp$, to couple the switching dynamics with the inclinations assumed by the potential. The panels (a), (b), (c) and (d) of the Fig.~\ref{PDF_RA} show $\Pe$ in function of time $t$, by changing the initial bias current values $i_0=\{$(a) 0, (b) 0.1, (c) 0.5, (d) 0.9$\}$. These data allow to explore the switching dynamics in correspondence of peculiar points of $\tau$ \emph{vs} $\omega$ curves (see solid lines lying on the $t$-$\omega$ planes and in Fig.~\ref{DynamicRA}), calculated for $\gamma=10^{-4}$. Setting $\omega=\wRA$, the resonance-like dynamics results in single-peack PDFs, centered around the MFPT values. This suggests that, in all the experiments, the phase particles tend to follow almost the same trajectory to escape from the initial metastable state. In particular, the panel (e) of Fig.~\ref{PDF_RA} shows the PDF calculated selecting $\omega=\wRAb=\{0.75, 0.95\}$, as a function of the normalized time $t/\Tp$. As already noted, setting $\omega=0.75$, that is in the first dynamic RA minimum, the particle tends to escape through the right barrier after almost one oscillation of the washboard potential, instead setting $\omega=0.95$, that is in the second dynamic RA minimum, the left-side escapes occur when $t\simeq 1.6\Tp$. These peacks show asymmetry and long tails. The asymmetry is more pronunced in low frequency data, due to the time that the washboard spend in the configurations supporting the escape events, that increase reducing its oscillation frequency. The insets in the panel (e) of Fig.~\ref{PDF_RA} show the PDFs plotted in function of the bias current $i_b(t)$, to compare, at least qualitatively, these data with the switching current probability $P(I_c)$, often studyed in the JJ framework. The shape of these single peack PDFs recall the termally activated switching current distribution in SGS systems (see Ref.~\onlinecite{Lee11,Cos12}), and their orientation depends to the washboard dynamics when the escape event occurs. 
The PDFs for frequencies within the dynamic RA minima are formed by single peacks, but tend to broaden moving from $\omega=\wRA$ (see panels (a) and (b) of the  Fig.~\ref{PDF_RA}). Far from these frequencies, that is when the $\tau$ values grow, the PDFs show multi-peacks structures, suggesting that the trajecotires followed by the phase particle in the various experiments get very spread. This occurs also for $\omega=0.81$ and $i_0=\{0.1, 0.5\}$, that is in correspondence of the narrow maximum ``intra-RA minima'' of $\tau$ data (see the panels (b) and (c)). The PDFs for high frequencies show high narrow peacks for $t=t_{max}$ indicating the particle inability to leave the metastable state, i.e. trapping events. In particular, for $i_0=0$ and $\omega=0.87$ almost $21\%$ of the experiments give entrapment, whereas for $i_0=0.1$ and $\omega=1.01$ this occurs in $\sim2\%$ of the experiments. Increasing $i_0$, the width of the peacks reduces, but the multi-peacks structures in high frequencies PDF are still evident as well as the trapping phenomena. In detail, selecting $i_0=5$ and $\omega=1.13$ the probability that the phase particle undergoes a trapping is $\sim4\%$, and for $i_0=9$ and $\omega=2.1$ is $\sim23\%$. The panels (a - f) of the Fig.~\ref{PDF_NES} show $\Pe$ in function of normalized time $t/\Tp$
obtained setting system and noise parameters in analogy with those setted in the panels (a - f) of the Fig.~\ref{NES}, that is: panel (a) $i_0=0$, $\omega=0.44$, panel (b) $i_0=0.1$, $\omega=0.44$, panel (c) $i_0=0.1$, $\omega=1.0$, panel (d) $i_0=0.5$ and $\omega=0.6$, panel (e) $i_0=0.5$, $\omega=1.08$ and panel (f) $i_0=0.9$, $\omega=1.18$. These PDF data allow to explore the NES effects described in the Fig.~\ref{NES} (solid curves reported on the $t$-$\omega$ planes of the panels of Fig.~\ref{PDF_NES}). The creation of NES maxima is due to the possibility that random fluctuations confine the particle inside the well also in the potential configurations good for escape events. In correspondence of the NES maxima, the PDF are composed by long regular sequences of peacks, with amplitude esponentially decreasing in time. For low potential inclination (see panels (a - c)) and $\gamma\leq10^{-1}$ these sequences are formed by two peacks per period, corresponding to a right- and left-side escapes, and every peacks have an almost triangular shape. Increasing the noise amplitude, the peacks tend to get less high but more large, melting in an almost trinagular large peacks for $\gamma\geq1$. The low frequency PDF (panels (a), (b) and (d)) show peacks spreading over almost 2, 1 and 0.5 periods $\Tp$ for $\gamma=1$, $10$ and $100$ respectively. Looking high frequencies PDF (panels (c), (e) and (f)) the width of these peack is twice, that is they spread over almost 4, 2 and 1 periods $\Tp$ for $\gamma=1$, $10$ and $100$ respectively. The PDF in the panel (d) doesn't show regularity, the potential is highly tilted and its frequency is not enough to generate long-living trajectories. The panels (e) and (f) regard highly tilted potentials too, but the frequency is high enough to trap the phase particle inside the initial well for long time generating interesting periodic peacks structures. Every period contains two asymmetric and very close peacks, a first intense and large due to right-side escapes and a second narrow connected with the left-side escapes.  
 
\section{Conclusions}

We explored the influence of thermal fluctuations on
the behavior of a ballistic graphene-based
Josephson junction in the short-junction regime. In particular, we
analyzed how random fluctuations affect the lifetime of the
superconductive state in an underdamped current-biased JJ. The
analysis was performed within the framework of the resistively and
capacitively shunted junction (RCSJ) model, using a proper
non-sinusoidal current-phase relation, characteristic of
graphene. Specifically we investigated the mean first passage time
(MFPT) of the phase particle, i.e. the phase difference across the
junction, initially placed in a minimum of the
tilted washboard-like potential. In particular, we
studied the MFPT as a function of different parameters of the
system and external perturbations, i.e. Gaussianly
distributed random fluctuations and periodical driving signal. We
found nonmonotonic behavior of the lifetime, $\tau$, of the
superconductive state as a function of the noise intensity $\gamma$,
driving frequency $\omega$ and fixing the initial value of the bias current $i_0$. 
These results indicate the presence of noise induced
phenomena, such as \emph{stochastic resonant activation} (RA) and
\emph{noise enhanced stability} (NES) with different features,
strongly depending on the initial value, $i_0$, of the bias current.
In particular, we observed ranges of parameters in which MFPT show
evidence of \emph{dynamic} and \emph{stochastic} RA, including a
multi-minimum RA effect in the low-noise-intensity regime. Finally,
we observed changes in the behaviour of MFPT, when the white noise
source is replaced by a coloured noise source with
different values of the correlation time $\tau_c$.\\
Our study provides information on the role played by
random (both thermal and correlated) fluctuations in the switching
dynamics from the superconductive state to the resistive one of a
graphene-based JJ. The results obtained can help to
better understand the role of fluctuations in the
electrodynamics of new generation graphene-based superconductive
devices, such as Josephson junctions, Josephson sensors, dc-SQUIDs
and gate-tunable phase
qubits, contributing to improve their performances.\\
In conclusion this work, which is well placed in the
framework of the nonequilibrium statistical mechanics, due to the
presence of an emerging material, such as graphene, with unique
electrical properties, presents relevant and interesting results
from several points of view.

\section*{Acknowledgements}

Authors acknowledge the financial support of Ministry of Education,
University, and Research of Italian Government (MIUR). They also
thank Dr. I. Hagym\'asi for fruitful discussions and useful
suggestions.


%

\end{document}